\documentclass[a4paper, draftcls, onecolumn, 12pt]{IEEEtran}
\usepackage[noadjust]{cite}
\usepackage{latexsym}
\usepackage[latin1]{inputenc}
\usepackage[T1]{fontenc}
\usepackage[dvips]{graphicx}
\usepackage{verbatim}
\usepackage{amsmath}
\usepackage{amsfonts}
\usepackage{amssymb}
\usepackage{exscale}
\usepackage{latexsym}
\usepackage{color}
\usepackage{bm}
\usepackage{stfloats}
\usepackage[nolists, nomarkers]{endfloat}
\usepackage[letterpaper, textwidth=6.5in, textheight=9in]{geometry}

\newcommand{\figw}{0.98\columnwidth}

\title{Cooperation in Carrier Sense Based Wireless Ad Hoc Networks - Part I: Reactive Schemes}
\author{\large{Andrea Munari$^{\ddag}$,~\IEEEmembership{Member,~IEEE}, \\Marco Levorato$^{*}$,~\IEEEmembership{Member,~IEEE}, Michele Zorzi$^{\dag}$,~\IEEEmembership{Fellow,~IEEE}}\\
\normalsize $\ddag$ German Aerospace Center (DLR), Inst. of Communications and Navigation, Oberpfaffenhofen, 82234 Weßling, Germany. \\
\normalsize $*$ Dept.\ of Electrical Engineering, Stanford University, Stanford, CA 94305 USA. \\
\normalsize $\dag$ Dept.\ of Information Engineering, University of Padova, via Gradenigo, 35131 Padova, Italy. \\
\normalsize e-mail: andrea.munari@dlr.de, levorato@stanford.edu,
zorzi@dei.unipd.it.
\thanks{Andrea Munari was with the Dept.\ of Information Engineering, University of Padova, via Gradenigo, 35131 Padova, Italy.
Part of this work has been presented at IEEE SECON 2009, Rome, Italy, June 22-26, 2009.}\vspace{-0.85cm}}

\date{}

\begin{document}

\setcounter{page}{0}
\maketitle
\thispagestyle{empty}

\vspace{-3mm}
\begin{abstract}
Cooperative techniques have been shown to significantly improve the performance of wireless systems. Despite being a mature technology in single communication link scenarios, their implementation in wider, and practical, networks poses several challenges which have not been fully identified and understood so far. In this two-part paper, the implementation of cooperative communications in non-centralized ad hoc networks with sensing-based channel access is extensively discussed. Both analysis and simulation are employed to provide a clear understanding of the mutual influence between the link layer contention mechanism and collaborative protocols. Part I of this work focuses on \emph{reactive} cooperation, in which relaying is triggered by packet delivery failure events, while Part II addresses \emph{proactive} approaches, preemptively initiated by the source based on channel state information. Results show that sensing-based channel access significantly hampers the effectiveness of cooperation by biasing the spatial distribution of available relays, and by inducing a level of spatial and temporal correlation of the interference that diminishes the diversity improvement on which cooperative gains are founded. Moreover, the efficiency reduction entailed by several practical protocol issues related to carrier sense multiple access which are typically neglected in the literature is thoroughly investigated.
\end{abstract}


\newpage
\section{Introduction} \label{sec:introduction}

Cooperative techniques have recently emerged as one of the most promising communication paradigms for ad hoc networks, thanks to their ability to dramatically improve the performance of wireless links \cite{Laneman-dec04,aaz_comm,aaz_comm2}. The basic idea that underpins these schemes is to let nodes in the proximity of an active source-destination pair, referred to as \emph{relays}, help data delivery by forwarding to the addressee redundancy on the message originally sent by the source. The receiver, by collecting fragments of information conveyed through the ideally independent source-to-destination and relay-to-destination channels, benefits from spatial diversity, so that the average performance of the link in terms of either achievable rate or outage probability improves.

While the ultimate capacity limit of the relay channel is still unknown~\cite{cover_relay}, there exists a wide range of well-studied cooperative techniques which provide sub-optimal performance~\cite{Lai06,capth}. Along this line, the simple Amplify and Forward (A\&F) approach~\cite{aef1,Laneman-dec04,far_af}, proposes relays to transmit an amplified version of the source's analog signal (plus noise), whereas with  Decode and Forward (D\&F)~\cite{Laneman-dec04,dec_raj}, Compress and Forward (C\&F)~\cite{Lai06,capth} and Coded Cooperation \cite{Hunter-feb06,elza_comm} the original message is decoded and processed by the relay before being forwarded to the destination.

Cooperation appears as a mature technology from the single data link perspective, yet its implementation in practical wireless networks poses several
challenges that need to be identified and addressed. In fact, although some recent works have started to investigate the impact that basic networking mechanisms such as medium access control may have on cooperative strategies \cite{MunariTWC,Levorato09} and, on the other hand, the influence exerted by relayed transmissions on the operations of a network \cite{MunariTWC,Badia10}, the complex nature of such interactions is far from being fully understood. Starting from these remarks, in this two-part paper we investigate the deployment of relaying techniques in large scale non-centralized wireless ad hoc networks with carrier sense-based channel contention. 

Among the set of relaying paradigms, the present work focuses on Coded Cooperation (CC), which implements a distributed Incremental Redundancy Hybrid Automatic Retransmission reQuest (IR HARQ) error correction scheme. According to the basic IR HARQ solution~\cite{harq_def}, a source node first encodes the data packet, and then sequentially transmits fragments of the obtained codeword in response to failure events reported by the destination through feedback packets, resulting in an extremely effective error control technique in time-varying channels~\cite{harq1,harq2}. Such an approach is extended in CC by having the transmission of additional redundancy performed by relays on behalf of the source, so as to increase channel diversity and achieve better performance. In this perspective, it is interesting to remark that not only does distributed IR HARQ provide good performance and robustness with respect to channel variations~\cite{spa_osv}, achieving the greatest diversity order compared to A\&F, D\&F and C\&F, but also it performs at its best when relay nodes are located in the proximity of the source \cite{Levorato09,Tomasin07}, condition that is often met under the carrier sensing access policy as will be extensively discussed in the remainder of this paper. The combination of these features, then, makes CC particularly apt to test the effectiveness of cooperation in CSMA-based scenarios.

In order to get a deep understanding of the interactions that exist between carrier sense-based access control and relaying and so as to draw broadly applicable conclusions, we follow the classification first introduced in~\cite{pro_rea} and consider two classes of solutions, namely \emph{reactive} and \emph{proactive} cooperative schemes. The former set, investigated in Part~I of this work, comprises approaches that trigger the intervention of relays only in the event of a missed decoding at the destination for the transmission performed by the source. Reactive cooperation is a viable, and relatively simple, solution for the implementation of distributed IR HARQ in ad hoc wireless networks, as it does not require complex coordination between the source and the available relays. In view of these properties, the study of reactive solutions is a first and fundamental step for the identification of the issues arising when cooperative communications are integrated in practical networking scenarios. 

Conversely, the companion paper~\cite{part2} focuses on \emph{proactive} relaying, which exploits channel state information to pre-emptively split a communication in two phases, transferring first the payload from the source to a third-party node and then letting such terminal send redundancy to the intended destination, with the aim of maximizing the sum rate of the link. The investigation of proactive solutions, which require a significant coordination overhead, provides an interesting parallel with state of the art single-link cooperative protocols and underlines their limitations in a networking environment.

With the aim of achieving a complete understanding of the dynamics of cooperation in a real-word network, we follow an approach based on both analysis and simulation, and we believe that both components are fundamental to the investigation presented in this paper. Indeed, on the one hand the analysis of simplified scenarios provides key insights on the influence of the contention mechanisms and interference from other active transmitters on the performance and statistics of cooperation. Nevertheless, accurate simulations of large networks capture a finer level of detail, highlighting many practical aspects that analysis cannot account for, and shed light on broader level statistics. 

Based on the results presented herein, we draw the following key observations:
\begin{itemize}
\item the presence of active transmitter-receiver pairs in the proximity of the nodes which are performing a cooperative communication may 
generate interference which is correlated in the time and space domain. Such an effect may reduce the cooperative gain, as it decreases channel diversity at the receiver of the link;
\item the carrier sense-based contention mechanism biases the spatial distribution of nodes available for cooperation, reducing the probability that a potential relay lies in the region providing optimal performance. As a result, the effectiveness of cooperation decreases;
\item practical issues, such as header decoding and receiver synchronization, hamper cooperative techniques in non-centralized ad hoc networks;
\item the extremely dynamic nature of interference in large and distributed CSMA-based networks, induced by the non-coordinated start and end of concurrent communications, significantly hinders the potential benefits brought by proactive relaying solutions.
\end{itemize}

The rest of the present Part~I is organized as follows. We start in Section~\ref{sec:systemModel} by introducing the system model and the notation considered throughout the paper. Then, Section~\ref{sec:dharqAnalysis} presents an analytical framework that investigates in simple topologies how CSMA biases some spatial distributions that are of paramount importance for the effectiveness of cooperation. Taking the cue from such analytical results, Section~\ref{sec:dharqDescription} introduces a simple protocol that extends basic CSMA to take advantage of coded cooperation, whose performance is thoroughly studied in Section~\ref{sec:dharqSims}. Finally, Section~\ref{sec:conclusions} concludes the paper.

\section{System Model and Notation} \label{sec:systemModel}

Carrier sense is a fundamental networking mechanism, widely implemented in present day ad hoc networks to control multiple access.\footnote{Throughout this paper we focus on plain carrier sense multiple access (CSMA), epitomized by the IEEE 802.11 DCF without channel negotiation \cite{802.11}, while an investigation including collision avoidance is left as part of future work.} According to CSMA, a terminal that has data to send picks a random backoff interval whose duration $n$, in slots, is uniformly drawn in $[1,2^{\textrm{CW}-1}]$, where $\textrm{CW} = \textrm{CW}_{start} + i$, CW$_{start}$ is a system parameter to handle congestion, $i = 0, \ldots, \textrm{SRL}-1$, and SRL is the Short Retry Limit, i.e., the maximum number of transmission attempts performed at the MAC layer before dropping a packet. During this interval, the node senses the aggregate power level on the channel. If the value exceeds the carrier sense threshold, the terminal stops the countdown and freezes the backoff until the medium is sensed idle again for at least a Distributed coordination function InterFrame Space (DIFS) period. On the other hand, once the backoff expires the node transmits its packet. If the destination succesfully decodes the payload, an acknowledgement (ACK) is sent to the source and the communication comes to an end.  Conversely, if the reception fails, no feedback is sent; the source increases its CW counter, and further attempts are performed after newly drawn random backoffs until either the packet is successful or the SRL is reached. 

The simple sensing mechanism implemented  by CSMA allows nodes to acquire information on their neighbors' activity, and is intended both to protect ongoing communications from interference and to block terminals whose transmissions would  fail with high probability. However, channel impairments such as noise, fading and shadowing as well as the well known hidden and exposed terminal problems~\cite{Tobagi75} may limit the effectiveness of this scheme. Nevertheless, since carrier sense based medium access strategies are widely implemented, they represent a significant  and insightful test environment for cooperative protocols.

In order to keep our scenario general and to have a framework compatible with many other theoretical studies, we model wireless links considering Rayleigh fading, so that the received power $\eta_{n_1,n_2}$ over the channel between two nodes $\bm N_1$ and $\bm N_2$ is an exponential random variable with mean $P \delta_{n_1,n_2}^{-\alpha}$, where $P$ is the transmission power, $\delta_{n_1,n_2}$ is the Euclidean distance between the two terminals and $\alpha$ is the path loss exponent. Moreover, we consider a channel capacity model for packet decoding, so that the failure probability in retrieving an $L$-bit payload for a transmission that starts at $t_s$ and lasts for $T$ seconds can be expressed as Pr$\left\{ \mathcal L < L\right\}$, where $\mathcal L$ is the number of decoded information bits given by:
\begin{equation}
 \mathcal L = \int_{t_s}^{t_s+T} \mathcal C(\gamma(t)) dt \,,
\label{eq:decodedInfoBits}
\end{equation}
and $C(\gamma(t)) = B \log_2(1 + \gamma(t))$ is the instantaneous link capacity between sender and receiver with bandwidth $B$ and Signal to Interference and Noise Ratio (SINR) $\gamma(t)$.

Throughout this paper, we denote scalars with regular font, whereas vectors are represented in bold. Moreover, the notation $\mathcal K^{x}$ indicates quantity $\mathcal K$ conditioned on the variable $x$. 

\section{Impact of CSMA on Cooperators Distribution} \label{sec:dharqAnalysis}

As discussed in Section~\ref{sec:introduction}, this paper focuses on relaying approaches that take place in case of a communication failure over a direct link so as to improve the chance of packet recovery, whereas the study of proactive schemes is addressed in a companion work \cite{part2}. Reactive cooperation techniques represent an interesting case study both because of the simple yet smart principle that they propose and because of the large deal of research that has been devoted to the analysis of their potential \cite{Laneman-dec04,aaz_comm,aaz_comm2}. Let us consider a data transfer between a source and a destination node. In case of a decoding failure, basic Automatic Repeat reQuest (ARQ) schemes would require the former to perform additional attempts delayed as per the rules of the underlying medium access control. However, if the communication has failed due to harsh channel conditions, e.g., fading, successive retransmissions are not likely to succeed unless realized after a time interval long enough for the channel to decorrelate. On the other hand, thanks to the broadcast nature of the wireless medium, other terminals may have received the packet even if not addressed to them. The reactive cooperative paradigm proposes that one of such nodes acts as relay by immediately sending redundancy to the destination on behalf of the source, so as to replace the time diversity that characterizes basic ARQ with spatial diversity.

\begin{figure}
\centering
\includegraphics[width=\figw]{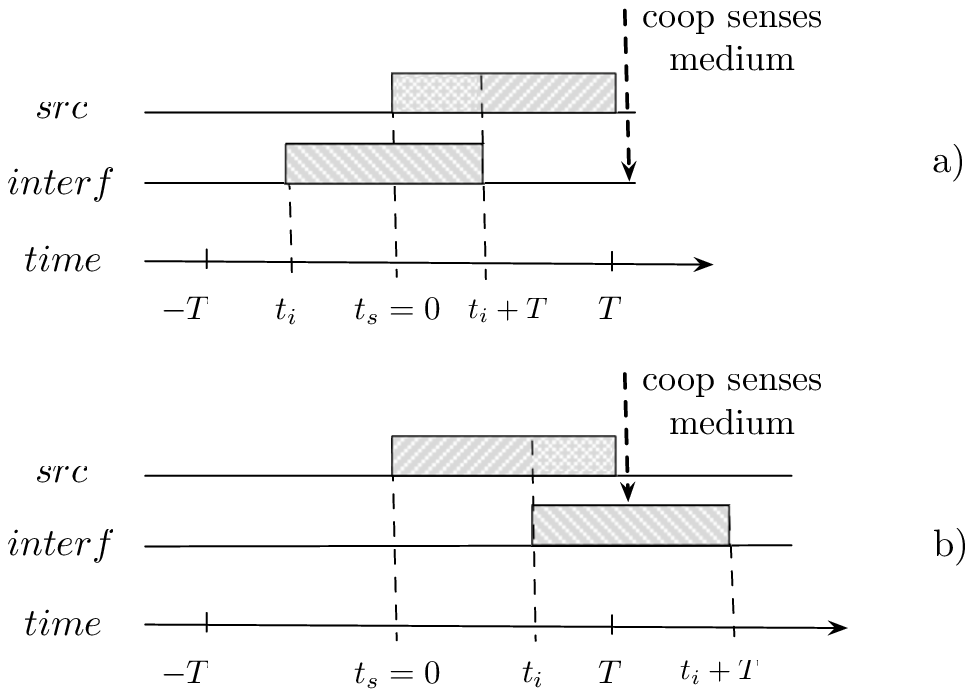}
\caption{Time diagram for the considered scenario. $\bm S$ accesses the channel at $t_s=0$, while $\bm I$ tries to send data at $t_i$. All communications last $T$ seconds, and the cooperating terminal senses the channel immediately after the end of $\bm S$'s transmission. a) $t_i < t_s$, b) $t_i \geq t_s$.}
\label{fig:rcAnalysis_timeDiagram}
\end{figure}

As a first step in the direction of understanding the mutual influence between carrier sense-based medium access and reactive cooperation, we develop an analytical framework to highlight some key interactions in simple topologies. To this aim, let us consider a scenario with a source $\bm S$, a destination $\bm D$, and an interfering terminal $\bm I$ deployed in a region $\mathcal A$ at positions $p_{s}=\{x_{s},y_{s}\}$, $p_{d}=\{x_{d},y_{d}\}$ and $p_{i}=\{x_{i},y_{i}\}$, respectively. All data transmissions in the network are performed at a fixed information bitrate $\rho_{data}$ and involve a payload of $L$ bits, thus lasting $T=L/\rho_{data}$ seconds. Furthermore, for the sake of mathematical tractability, we assume that fading coefficients remain constant for the whole duration of a data packet.\footnote{The assumption of constant fading coefficients will be relaxed in the simulation studies presented in this work by accurately modeling the time-correlated nature of the wireless channel, see Section~\ref{sec:dharqSims}. Moreover, note that the link capacity $C(t)$ reported in (\ref{eq:decodedInfoBits}) may vary during a communication even in the presence of a constant fading, due to changes in the perceived aggregate interference level.}
Without loss of generality, as reported in Fig.~\ref{fig:rcAnalysis_timeDiagram}, $\bm S$ accesses the channel at time $t_s = 0$, while the birth time $t_i$ of the interfering communication is distributed in $[-T, T]$ according to a generic probability density function $f_{t_i}$. Recalling the system model described in Section~\ref{sec:systemModel}, the probability that a node $\bm N_2$ senses the medium idle given that terminal $\bm N_1$ is transmitting, i.e., the carrier sense constraint, is given by:
\begin{equation}
 \mathcal F (p_{n_1}, p_{n_2}) \!=\! \textrm{Pr}\left\{  \eta_{n_1,n_2} + N \!<\! \Lambda \right\} \!=\! 1 -
  e^{-\frac{\Lambda - N}{P \delta_{n_1,n_2}^{-\alpha}}}  \,,
\label{eq:csConstraint}
\end{equation}
where $N$ is the noise floor and $\Lambda$ is the carrier sense threshold.

We begin our study by computing, for a given topology configuration $\bm p = \{p_{s}, p_{d}, p_{i} \}$, the probability $\mathcal I(\bm p)$ that the interferer induces an outage at the destination when carrier sense is employed as the medium access policy, thus triggering the need for cooperation. In order for this to happen, two conditions have to be met. In the first place, both $\bm S$ and $\bm I$ have to be granted access to the channel. This implies that either the source, for $t_i \leq 0$, or the interferer, for $t_i > 0$, are required to sense a power level on the medium below the carrier sense threshold despite the other node's ongoing transmission. Secondly, the interference level at $\bm D$ during the overlapping part of the two communications has to be high enough to prevent successful decoding of the payload. The two events are
statistically independent, since the former only involves the $\bm I$-$\bm S$ link, whereas the latter depends on the $\bm S$-$\bm D$ and $\bm I$-$\bm D$ channels. Therefore, conditioning on the birth time $t_i$ of the interfering transmission, and recalling the CSMA
constraint reported in (\ref{eq:csConstraint}), the outage probability can be factorized as:
\begin{equation}
\mathcal I^{t_i}(\bm p) = \mathcal F(p_{i},p_{s}) \, \mathcal O^{t_{i}}(\bm p)  \,,
\label{eq:factorizationI}
\end{equation}
where $\mathcal O^{t_i} (\bm p) = \textrm{Pr}\{ \mathcal L^{t_i} < L\}$ is the probability
of decoding less than $L$ information bits at $\bm D$ conditioned on $t_i$.
Notice that for both $t_i \leq t_s$ and $t_i > t_s$, $T - |t_i|$ seconds of $\bm S$'s payload
are affected by $\bm I$'s transmission, whereas $|t_i|$ seconds are interference-free.
Thus, applying (\ref{eq:decodedInfoBits}), and defining as $\bm \eta = \{ \eta_{s,d}, \eta_{i,d}
\}$ the vector of powers received at $\bm D$, the number of retrieved information bits at the destination
evaluates to:
\begin{equation}
 \mathcal{L}^{t_{i}}(\bm \eta) = |t_i| \, \mathcal C\left( \frac{\eta_{s,d}}{N}\right)
 + (T - |t_i|) \, \mathcal C \left( \frac{\eta_{s,d}}{N + \eta_{i,d}} \right) \,.
\label{eq:decodedInfoBits_ti}
\end{equation}
The explicit derivation of $\mathcal O^{t_{i}}(\bm p)$, though conceptually simple, is
rather involved, and is reported in Appendix~\ref{appendixDHARQ_interf}. Taking advantage of
the result reported in (\ref{eq:outageProb_ti}), we can compute the sought outage probability by
numerical evaluation of:
\begin{equation}
 \mathcal I(\bm p) = \int_{-T}^{T} \mathcal{I}^{t_i}(\bm p) \, f_{t_i}(t_i) \;dt_i\;.
 \label{eq:interfDistribution}
\end{equation}

\begin{figure}
\centering
\includegraphics[width=\figw]{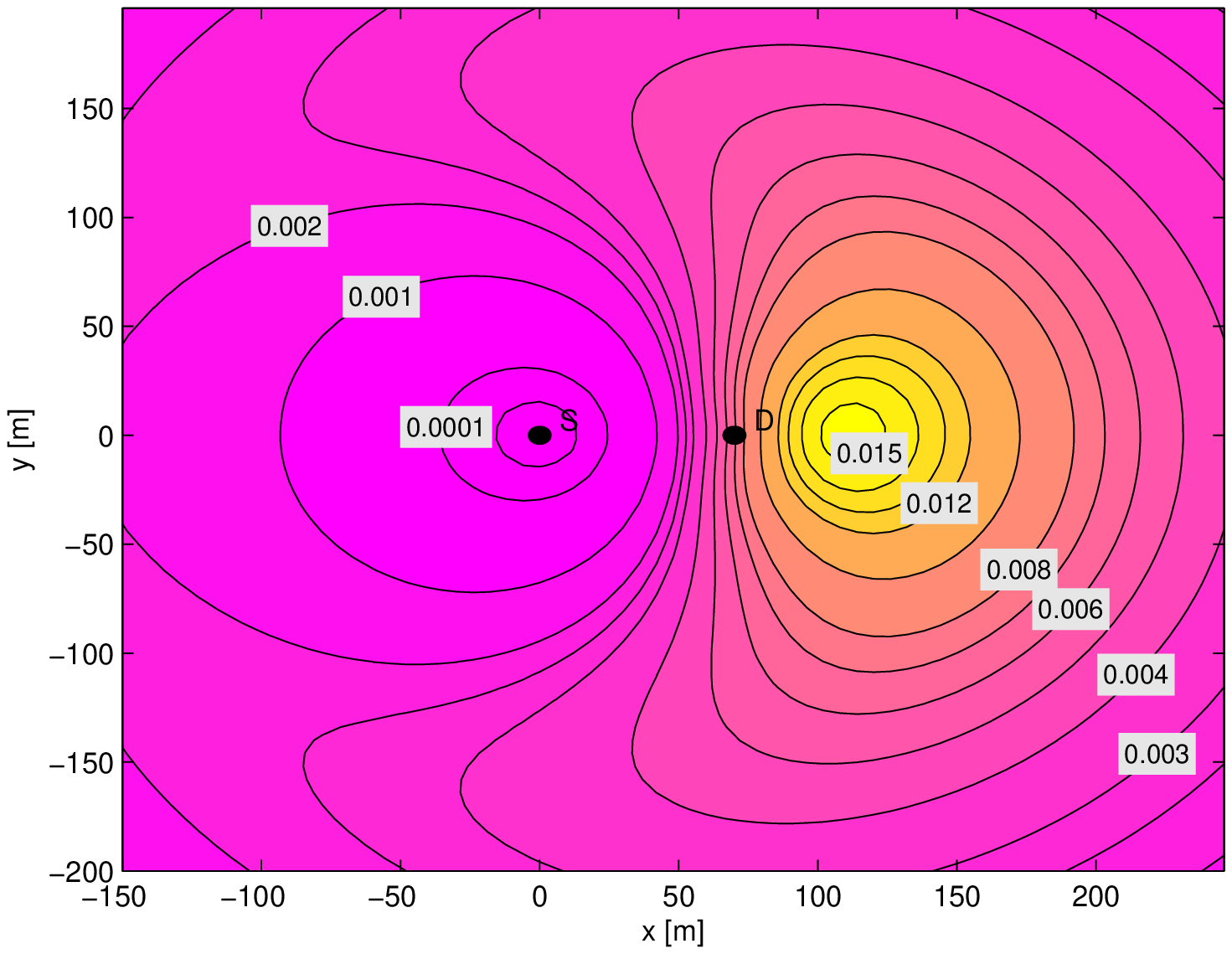}
\caption{Probability that an interfering node located at $p_{i} = \{x,y\}$ causes an outage event at $\bm
D$. $p_s = \{0,0\}$, $p_d = \{60,0\}$.}
\label{fig:rcAnalysis_InterfDistrib}
\end{figure}

While the developed framework applies to any distribution for the birth time of the interfering communication, we study in greater detail the uniform case, i.e., $f_{t_i}(u)=1/2T,\, u\in[-T,T]$, in view of both its generality and tractability. In particular, Fig.~\ref{fig:rcAnalysis_InterfDistrib} depicts $\mathcal I(\bm p)$ under this assumption for different positions of $\bm I$, when the source and the destination are placed at $p_s = \{0,0\}$ and $p_d = \{60,0\}$, respectively, with the coordinates expressed in meters, and when all the other system parameters are set according to Tab.~\ref{tab:parameters}. It is possible to observe that the simple CSMA policy induces a strong
bias in the spatial distribution of interferers when cooperation is needed. Indeed, the region in which
$\bm I$ is most likely to induce a failure lies along the line connecting $\bm S$ and $\bm D$,
right behind the destination. This result is rather intuitive, since such positions are at the
same time geographically close to $\bm D$, so that $\bm I$ can heavily interfere with the
reception, and sufficiently far from $\bm S$, so that the carrier sense constraint can be met with
high probability.

Even more interesting for our discussion is the fact that such an interferer distribution
significantly influences the availability and the position of relay nodes, with potentially
detrimental effects on reactive cooperation. To investigate this aspect, let us consider a terminal $\bm C$ located
at $p_c = \{ x_c, y_c \}$, and assume that it may act as helper for the $\bm S$-$\bm D$ link if two
conditions are satisfied: i) it has decoded the payload sent by the source despite the interfering
transmission which induced an outage at the destination; and ii) it senses the medium idle at time
$t=T$ and is therefore allowed to send redundancy to $\bm D$. Both requirements clearly depend on the position of $\bm I$, which, in turn, varies with the birth time of the interfering communication, as discussed earlier and as expressed by (\ref{eq:factorizationI}). Therefore, defining $\bm p' = \{ p_s, p_d, p_c \}$ as a vector describing the topology under consideration, the spatial distribution of cooperators can be determined by first evaluating the probability $\mathcal H^{p_{i}, t_{i}}(\bm p')$ that $\bm C$ meets i) and ii) conditioned on $p_{i}$ and $t_{i}$, and by then averaging this distribution over the interference conditions. In particular, the probability $\mathcal H^{t_{i}}(\bm p')$ that $\bm C$ is available given that the interfering transmission starts at $t_{i}$ can be computed as:
\begin{equation}
\mathcal H^{t_i}(\bm p') = \frac{\int_{\mathcal A} \mathcal H^{p_i,t_i}(\bm p')
\, \mathcal I^{t_i}(\bm p) \;d p_i}{\int_{\mathcal A} \mathcal I^{t_i}(\bm p) \;
d p_i} \;,
\end{equation}
with  $\mathcal I^{t_i}(\bm p)$ given by (\ref{eq:factorizationI}). In turn, the unconditional probability $\mathcal H(\bm p')$, i.e., the sought cooperators distribution, is obtained as:
\begin{equation}
 \mathcal H(\bm p') = \int_{-T}^{T} \mathcal H^{t_{i}}(\bm p') \, f_{t_i}(t_i) \,dt_{i} \;.
\label{eq:H_sum}
\end{equation}

\begin{table}[b]
  \begin{center}
  \caption{Parameters used in our studies}
  \label{tab:parameters}
  \begin{tabular}{|c|c|}
  \hline
  Transmission power (dBm) & 10 \\
  \hline
  Noise Floor (dBm) & -102 \\
  \hline
  Detection threshold (dBm) & -96\\
  \hline
  Path loss exponent, $\alpha$ & 3.5\\
  \hline
  Maximum Doppler frequency, $f_d$ (Hz) & 11.1\\
  \hline
  Carrier Frequency (GHz) & 2.4\\
  \hline
  Bandwidth, $B$ (MHz) & 1\\
  \hline
  Headers and control pkt information bitrate, $\rho_{ctrl}$ (Mbps) & 0.532 \\
  \hline
  Data pkt information bitrate, $\rho_{data}$ (Mbps) & 2.1 \\
  \hline
  CS threshold, $\Lambda$ (dBm) & -100\\
  \hline
  CS threshold for relay contention, $\Lambda_{rel}$ (dBm) & -100\\
  \hline
  Slot, DIFS, SIFS duration ($\mu$s) & 10, 128, 10\\
  \hline
  Initial contention window index, CW$_{start}$ & 5 \\
  \hline
  Number of slots used for relay contention, $\textrm{CW}_{rel}$ (slot) & 16 \\
  \hline
  SRL - cooperative, basic protocols & 4, 5 \\
  \hline
  Levels of quantization for $\gamma_{s,d}$ in the NACK frame & 8 \\
  \hline
  Interference margin, $\varepsilon$ & 0.06 \\
  \hline
  DATA header (bit) & 112 \\
  \hline
  Payload (bit) & 5000 \\
  \hline
  ACK/NACK (bit) & 112 \\
  \hline
  \end{tabular}
  \end{center}
   \vspace{-10pt}
\end{table}

\begin{figure}
\centering
\includegraphics[width=\figw]{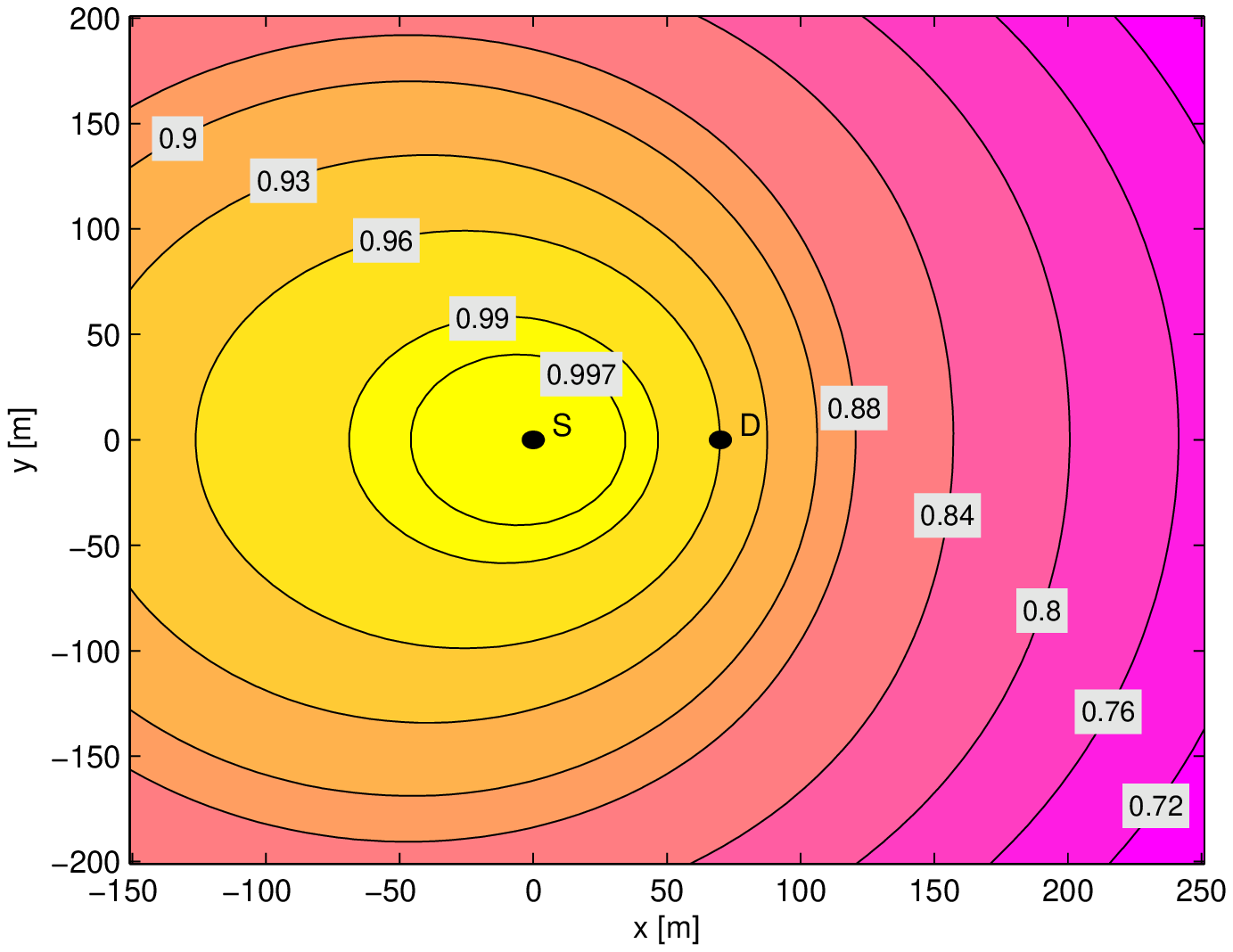}
\caption{Probability that a node located at $p_{c} = \{x,y\}$ is available for cooperation given that the
interfering transmission has already come to an end.}
\label{fig:rcAnalysis_CoopDistribFirstHalf}
\end{figure}

In order to derive $\mathcal H^{p_i,t_i}(\bm p')$, it is useful to distinguish the cases
for $t_i \leq 0$ and $t_i > 0$. In the former condition, as reported in Fig.~\ref{fig:rcAnalysis_timeDiagram}, $\bm I$'s transmission has
already come to an end when $\bm C$ potentially senses the medium for channel access.
Therefore, requirement i) is always met, and the only constraint for the node to act as
cooperator is the decoding of the payload sent by $\bm S$ given the interference level generated
by $\bm I$. This event, in turn, can be analyzed following the same procedures we discussed for
the study of the outage at $\bm D$, and we immediately get:
\begin{equation}
 \mathcal H^{p_i, t_i}_-(\bm p') = 1 - \mathcal O^{t_i}(p_s,p_c,p_i)\;,
\end{equation}
where the subscript  ``-'' of $\mathcal H^{p_i,t_i}_-$ indicates that we are considering values of $t_i
\leq 0$, and $\mathcal O^{t_i}(p_s,p_c,p_i)$ can be obtained by computing
(\ref{eq:outageProb_ti}), reported in Appendix~\ref{appendixDHARQ_interf}, on the topology $\{p_s,
p_c, p_i \}$. Conversely, for $t_i > 0$, $\bm I$ is still transmitting when the relay candidate
checks the power level on the medium, and thus the interference level perceived at $\bm C$, i.e.,
$\eta_{i,c}$, determines both the payload decoding and the carrier sensing events. This implies
that, unlike what has been done in the first part of our discussion, the conditional probability
$\mathcal H^{p_i,t_i}_+(\bm p')$ cannot be factorized to take into account separately
requirements i) and ii).\footnote{In accordance to the notation introduced earlier, the subscript ``+'' in $\mathcal H_{+}^{p_{i},t_{i}}(\bm p')$ specifies the fact that we are dealing with interfering transmissions characterized by $t_i > 0$.} Its derivation, consequently, becomes rather involved, and for
the sake of clarity is postponed to Appendix~\ref{appendixDHARQ_coop}. In this
section, instead, we focus on the discussion of the cooperator distribution obtained by applying
the presented analytical approach to a topology with source and destination 60 m apart, located once again at $p_s =
\{0,0\}$ and $p_d = \{60, 0\}$ respectively, and assuming a uniform distribution for the birth time of the interfering transmission, i.e., $f_{t_i}(u) = 1/2T,\, u\in[-T,T]$.

As a starting point, Fig.~\ref{fig:rcAnalysis_CoopDistribFirstHalf} reports the probability that a
terminal is available as relay conditioned on having the interfering transmission start before
$t_s$, i.e., (\ref{eq:H_sum}) restricted to the integration domain $[-T,0)$. In this situation, as discussed, the only requirement $\bm C$ has to meet is to have decoded the payload sent by the source. From the plot
it is apparent how such a constraint is satisfied with higher probability when the
terminal is located in the proximity of $\bm S$. This result stems from the interferer
distribution induced by CSMA and presented in Fig.~\ref{fig:rcAnalysis_InterfDistrib}, which
hampers payload retrieval especially for terminals which are close to $\bm D$.

\begin{figure}
\centering
\includegraphics[width=\figw]{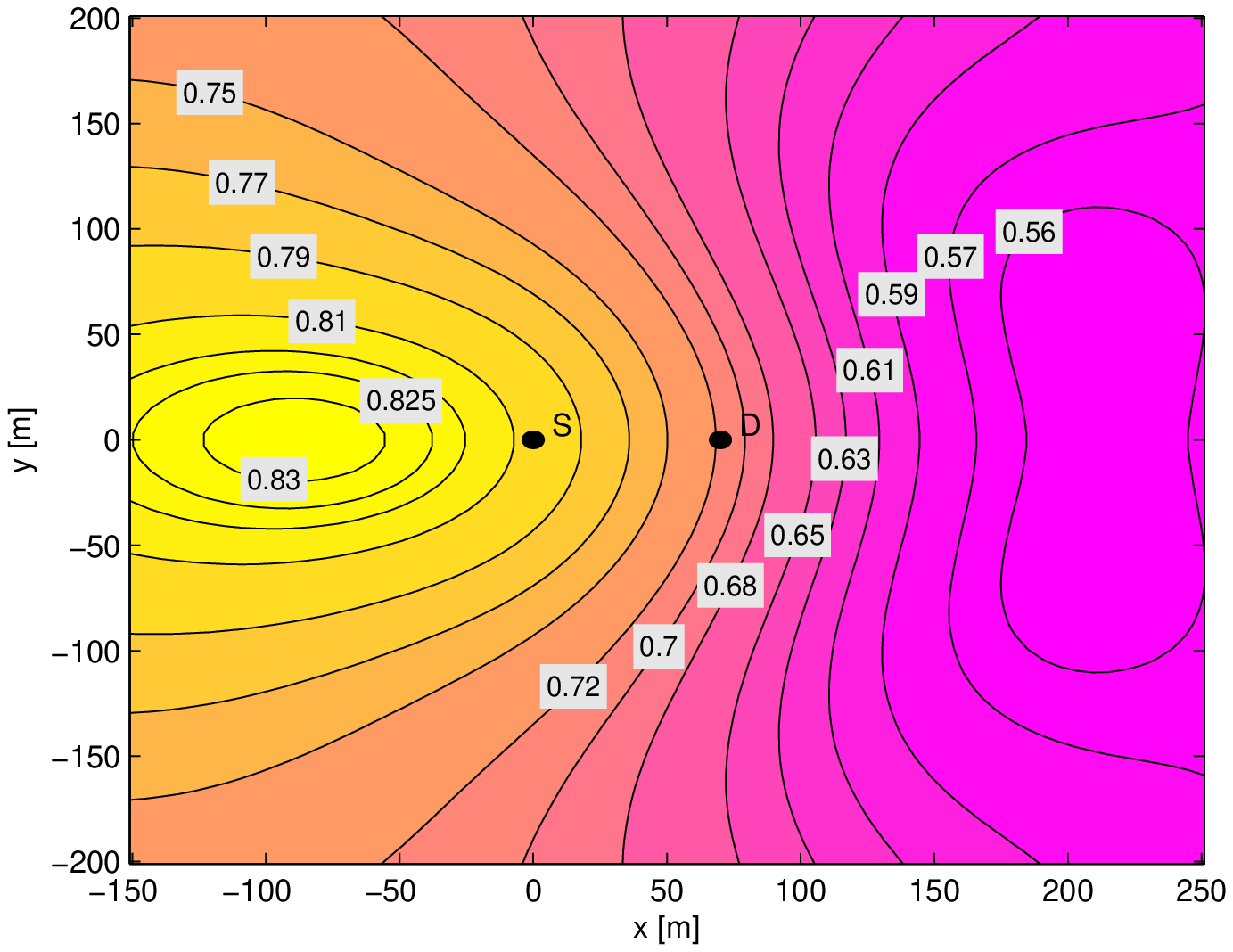}
\caption{Probability that a node located at $p_{c} = \{x,y\}$ is available for cooperation given that the
interfering transmission is still active.}
\label{fig:rcAnalysis_CoopDistribSecondHalf}
\end{figure}

The spatial distribution of cooperating nodes becomes even more biased when $\bm I$'s transmission
is still active at the end of the $\bm S$-$\bm D$ data exchange, as shown
in Fig.~\ref{fig:rcAnalysis_CoopDistribSecondHalf}, which depicts (\ref{eq:H_sum}) computed on the
restricted integration domain $[0, T]$. In such a case, relaying may not take place even when the packet sent by $\bm S$ has been correctly received at $\bm C$, since the terminal may refrain from cooperating due to the carrier sense constraint, so as not to harm the communication being performed by $\bm I$. The impact of this condition  on the availability of relays is significant even in the minimal topology considered here. Indeed, when only packet decoding from $\bm S$ is requested, finding a cooperator in the proximity of the destination is only approximately 5\% less probable than having it close to the source (see Fig.~\ref{fig:rcAnalysis_CoopDistribFirstHalf}). Conversely, such a gap more than doubles when $\bm C$ is also subject to CSMA (see Fig.~\ref{fig:rcAnalysis_CoopDistribSecondHalf}).

\begin{figure}
\centering
\includegraphics[width=\figw]{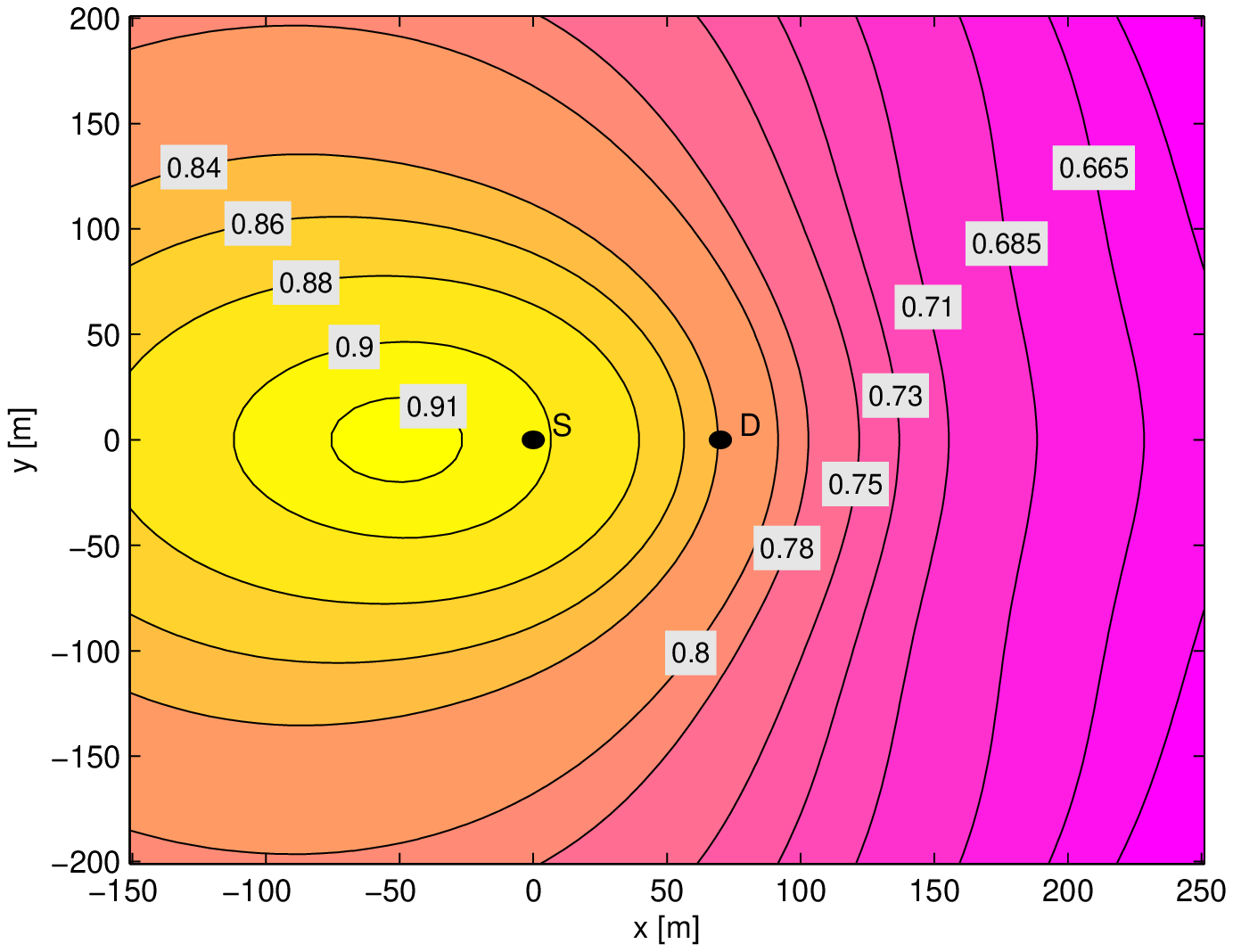}
\caption{Average probability that a node in $p_{c} = \{x,y\}$ is available for cooperation.}
\label{fig:rcAnalysis_CoopDistribAverage}
\end{figure}

Finally, Fig~\ref{fig:rcAnalysis_CoopDistribAverage} reports the spatial distribution $\mathcal
H(\bm p')$ of potential cooperators averaged over the whole birth time interval of the
interfering communication. As a first remark, we notice that nodes available for cooperation are concentrated close to $\bm S$, and thus offer a limited advancement towards the destination. As will be discussed in Section \ref{sec:dharqSims}, and as confirmed by further analytical results not reported here due to space constraints, this effect increases as the distance between $\bm S$ and $\bm D$ becomes larger. This stems from the fact that the  closer the source to the destination, the more the area reserved by the carrier sense mechanism to the ongoing transmission extends over the addressee's neighborhood, improving the decoding capabilities of relay candidates surrounding $\bm D$. However, cooperation is more likely to be required when the distance between the communicating terminals is large, due to the poorer link quality induced by the higher path loss.

We can therefore already infer from this simple analysis that CSMA has an unfavorable impact on cooperative schemes, since when relaying is needed, cooperators are likely to be far from the destination they want to reach.

\section{Distributed Hybrid ARQ in CSMA-based networks}
\label{sec:dharqDescription}

The analytical framework developed in Section~\ref{sec:dharqAnalysis} has highlighted how CSMA influences the geographical position of nodes available for reactive cooperation, concentrating them in the proximity of the source rather than close to the destination of a wireless link. This result, though obtained in simple topologies, offers important insights on the design of relaying protocols in carrier sense-based networks.

In the first place, we can infer that the biased distribution of relay candidates discourages the use of solutions that implement some form of \emph{cooperative routing}~\cite{coop_rout}. These approaches, indeed, exploit cooperation to build opportunistic paths toward the final destination, under the rationale of finding relays that provide a significant geographical advancement towards the addressee of the payload. Simulations of a simple cooperative routing protocol in a detailed network scenario, whose results are not reported here due to space constraints, have confirmed this intuition.

From this viewpoint, instead, distributed Incremental Redundancy Hybrid Automatic Repeat reQuest (IR HARQ) \cite{Zhao-jan05} appears as an ideal solution to reap the full benefits of cooperation when CSMA is used as medium access policy. According to these schemes, also known as \emph{coded cooperation} protocols, the destination collects fragments of a codeword obtained encoding the payload and received from the source and from relay nodes. Not only is coded cooperation able to exploit the greatest diversity order compared to all other approaches, but also it is likely to undergo a smaller performance reduction as a consequence of the worse position of relays, achieving its best results exactly when the cooperators are concentrated towards the source \cite{Levorato09,Tomasin07}. 

Starting from these remarks, in this section we present a medium access protocol named DHARQ that extends plain CSMA to take advantage of distributed HARQ of type II in {non-infrastructured} ad hoc networks. 
As in the non cooperative case, according to DHARQ a source $\bm S$ accesses the channel after the carrier sense-based backoff procedure described in Section~\ref{sec:systemModel}. Let us assume that no information on the channel connecting $\bm S$ to its addressee is available, so that the node sends its payload to the destination $\bm D$ at a fixed information bitrate $\rho_{data}$. If the transmission succeeds, an ACK at rate $\rho_{ctrl} < \rho_{data}$ is sent by the addressee, and the communication comes to an end. On the contrary, if the reception fails, two conditions may have occurred at $\bm D$: i) the header has not been successfully decoded, or ii) the header has been decoded, yet the payload was corrupted.\footnote{Note that condition ii) may occur since header and payload are typically independently encoded and have a separate CRC.} The former case may happen if the receiver's radio is synchronized to another ongoing transmission, e.g., due to the hidden terminal problem, or in the presence of harsh channel and interference conditions. In such a situation, the destination is not even aware of the attempt performed by $\bm S$. Therefore, no feedback can be sent, and the source relies on the basic CSMA approach for retransmissions, performing other attempts after suitable backoff intervals, until either the payload is successfully delivered or the SRL is reached. Conversely, if condition ii) occurs, the destination caches the corrupted version of the payload and transmits a Not ACKnowledgement (NACK) packet at rate $\rho_{ctrl}$, explicitly asking for an immediate retransmission from a relay node. Not only does the NACK frame trigger a hybrid ARQ procedure, but also it includes an indicator on the quality of the reception at $\bm D$. This is achieved by means of a $k$ bit field in the frame that contains a quantized version $\tilde\gamma_{s,d}$ of the average SINR perceived by the node during data reception. Any terminal $\bm C$ that receives the NACK packet and that has correctly decoded the payload transmitted by the source performs adaptive rate selection by computing the maximum information bitrate $\rho^{c}_{rel}$ it can use to perform a successful coded cooperative retransmission. To this aim, $\bm C$ evaluates the number $\mathcal L$ of decodable information bits at the destination after a relaying phase assuming constant SINRs during packet receptions and applying (\ref{eq:decodedInfoBits}):
\begin{equation}
\mathcal{L} = T_{s,d} \, \mathcal C(\tilde \gamma_{s,d}) + T_{c,d} \, \mathcal C(\gamma_{c,d}) \,,
\label{eq:sourceAndRelay}
\end{equation}
where $T_{s,d} = L/\rho_{data}$, $T_{c,d} = L/\rho^c_{rel}$, and $\gamma_{c,d}$ is the average SINR perceived by the relay candidate during the reception of the NACK packet. Note that (\ref{eq:sourceAndRelay}) describes the number of decodable information bits at $\bm D$ if node $\bm C$ acts as cooperator considering an estimate of the relay-destination channel $\gamma_{c,d}$ obtained at the former rather than at the latter terminal. However, even if the fading coefficient for the $\bm C$-$\bm D$ link is symmetric, the two overall SINRs may be significantly different, as the relay node is likely to experience a lower interference level, since it is typically more protected by the carrier sense region induced by the source's transmission. In order to cope with this problem, DHARQ selects the cooperative bitrate in a conservative fashion, assuming that  $L(1 + \varepsilon)$ information bits have to be decoded at $\bm D$. Thus, node $\bm C$ computes $\rho^c_{rel}$ from (\ref{eq:sourceAndRelay}) so as to satisfy the condition $\mathcal L \geq L(1 + \varepsilon)$, obtaining:
\begin{equation}
\label{ratecoop}
\rho^c_{rel} = \frac{\rho_{data} \, \mathcal C(\gamma_{c,d})}{(1 + \varepsilon)\, \rho_{data} - \mathcal C(\tilde \gamma_{s,d})} \,.
\end{equation}

Once the maximum sustainable rate has been determined, the terminal compares it to the one of the direct $\bm S$-$\bm D$ link. If $\rho^c_{rel} \leq \rho_{data}$, the node gives up the procedure and goes back to its activity, since its contribution would last longer than a retransmission attempt made by the source, strongly reducing the advantages of cooperation. On the other hand, if $\rho^c_{rel} > \rho_{data}$, the terminal enters a distributed contention to act as the relay by starting a random backoff whose duration in slots is uniformly drawn in the interval $[1,\textrm{CW}_{rel}]$. During this period, the candidate performs carrier sense. If the aggregate power level on the channel $P_{agg}$ exceeds a given threshold $\Lambda_{rel}$, the node assumes that another potential cooperator has chosen a shorter backoff window, and abandons the contention. If, instead, the countdown expires with an idle medium, the terminal transmits incremental redundancy at the rate $\rho^c_{rel}$ it has computed, and then goes back to its own activity. Let us stress that $\Lambda_{rel}$ is a fundamental parameter for our cooperative approach. In fact, while a conservative threshold would limit the number of nodes that can act as cooperators, an aggressive choice may cause unexpected and harmful interference to other neighboring transmissions, reducing the spatial reuse in the network.
A more in depth discussion on these tradeoffs will be provided in Section~\ref{sec:dharqSims}.

If the destination successfully decodes the data, it replies with an ACK at rate $\rho_{ctrl}$ addressed to $\bm S$. Otherwise, no feedback is sent. The lack of an explicit acknowledgment is interpreted by the source as a failure of the cooperative mechanism, and triggers successive attempts following the procedure described so far, until the packet is successfully received or the SRL is reached. 

Let us now make some remarks on the described protocol. DHARQ extends in a simple way the plain IEEE 802.11 DCF so as to take advantage of coded cooperation. From this viewpoint, the relay selection procedure that we propose has been designed trying to adhere to the carrier sense paradigm as well as to the completely distributed nature of non-infrastructured ad hoc networks.
Also, the presented medium access scheme triggers cooperation by sending a NACK only when it is actually needed, i.e., when the destination has failed to decode the data packet but has been able to cache a corrupted version of the payload. On the one hand, this approach introduces additional overhead. Nevertheless, the transmission of NACK frames prevents cooperative transmissions that could not be exploited, e.g., because $\bm D$ cannot perform packet combining, and that would only result in additional and harmful interference. Moreover, the provided feedback makes it possible to perform rate adaptation and to identify the relay node among terminals that are not involved in other communications and that are experiencing channel conditions with $\bm D$ that satisfy a minimum quality level, i.e., good enough to decode the NACK.

In this context, it is also worth noticing that the proposed solution does not apply any criterion for identifying cooperating nodes, but rather randomly chooses one terminal from a set of available candidates. Relay selection procedures have been widely studied in the literature, both from a theoretical perspective, finding optimal solutions \cite{Nosratinia07}, and from a more practical angle, identifying suboptimal or heuristics strategies capable of working well in realistic networking scenarios \cite{Jing09}. However, we have chosen not to take advantage of such approaches in DHARQ in order to keep its implementation as simple as possible and to highlight the impact of carrier sense on the efficiency of the plain cooperative paradigm. Moreover, the effects of different relay selection strategies will be discussed in Section~\ref{sec:dharqSims} by considering an idealized scheme that bounds the performance of the proposed protocol always resorting to the best possible candidate, i.e., the node that minimizes the overall transmission time. 

Finally, also notice that, for the sake of mathematical tractability, the framework developed in Section~\ref{sec:dharqAnalysis} did not consider NACK packets, only asking nodes to retrieve the payload from the source and to sense the medium idle in order to act as cooperators. We remark that this simplification does not significantly affect the insights presented on the cooperators distribution, as frames sent at rate $\rho_{ctrl}$ are decoded with high probability in the whole surroundings of $\bm S$ and $\bm D$. Therefore, the conclusions inferred on the impact of carrier sense-based medium access on cooperation also apply to the DHARQ protocol described in this section.
\section{Simulation Results for Reactive Cooperation}
\label{sec:dharqSims}

In order to test the effectiveness of reactive cooperation in large scale CSMA-based environments, extensive Omnet++ \cite{Varga01} simulations have been performed. In our study we have considered several configurations, analyzing different system cardinalities, playground sizes and networking parameters. However, due to space constraints and in view of the similar exhibited key trends, we focus our discussion on a specific and insightful case, given by networks composed of 35 static nodes spread over a {300$\,m \times 300\,m$} area. Single hop flows have been analyzed, with each terminal generating Poisson traffic addressed to its neighbors with intensity $\lambda$ (kbit/s/node). Such a configuration provides enough spatial separation to allow the presence of simultaneous communications within the network and is thus representative to test protocols under harsh channel contention, interference and hidden terminal conditions. The wireless environment, in line with the framework developed in Section~\ref{sec:dharqAnalysis}, is subject to frequency flat Rayleigh fading. In addition, our simulations take into account the temporal coherence of the wireless medium, which is modeled according to Jakes' approach for land mobile fading \cite{Jakes93}, so that the correlation between two instances of the same channel spaced in time by $\tau$ seconds is given by $J_0(2\pi f_d \tau)$, where $J_0(\cdot)$ is the zero-order Bessel function of the first kind and $f_d$ is the maximum Doppler frequency for the environment under consideration.
The information bitrate $\rho_{data}$ for the transmission of data packets has been chosen to guarantee a success probability of 0.99 at a distance of 60$\,m$ with 4 successive independent attempts in the absence of interference, and has been set to 2.1 Mb/s. In turn, the information bitrate $\rho_{ctrl}$ for headers and control packets has been set to 0.53 Mb/s, so as to achieve a decoding probability of 0.95 with a single transmission.

In our simulations, we have analyzed the performance of DHARQ and of a plain CSMA strategy. In order to have a fair comparison between these schemes, we have tuned the protocol parameters so as to achieve a similar reliability. In particular, a Packet Delivery Ratio of approximately {95\%} has been obtained by setting the SRL to 5 for CSMA, whereas 4 successive attempts were enough for DHARQ. The standard set of network and protocol parameters used in our studies are reported in Tab. \ref{tab:parameters}. All the results presented in this paper have been obtained averaging the outcome of 50 independent simulations, so that the 95\% confidence intervals, although not shown for readability, never exceed 3\% of the estimated value.

As a starting point for our studies, we have tested CSMA and DHARQ in networks composed by a single source node surrounded by a few neighboring terminals. This setting reproduces the average topology for a link in the reference network described earlier, yet it avoids the harmful effects induced by concurrent transmissions. The outcome of these investigations, not reported here due to space constraints, supported the trends predicted by many theoretical studies, e.g., \cite{Levorato09}, with distributed hybrid ARQ exhibiting a throughput that more than doubled the one achieved by plain CSMA.

\begin{figure}
\centering
\includegraphics[width=\figw]{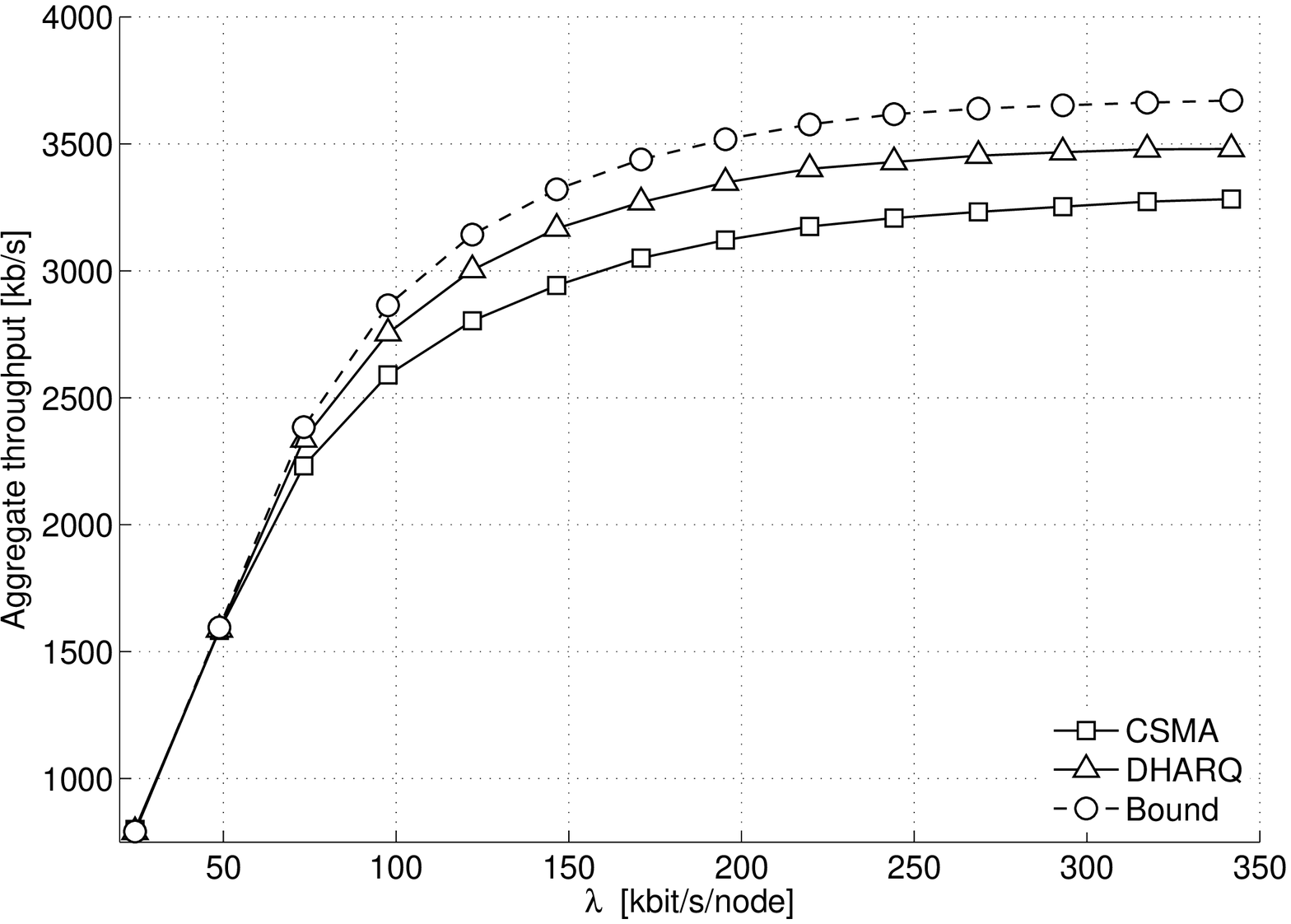}
\caption{Aggregate throughput vs nominal load.}
\label{fig:rcSims_truBound}
\end{figure}

However, when tested in more articulated and realistic scenarios reactive cooperation has not been able to reproduce similar gains. This is apparent in Fig.~\ref{fig:rcSims_truBound}, where the aggregate throughput against the nominal load $\lambda$ in the considered 35-node topologies is shown. In this case, DHARQ outperforms basic CSMA by less than {10\%} at saturation. This observation is further corroborated by the dashed curve reported in the plot, which describes the behavior of a bound for the class of protocols that implement the reactive cooperative paradigm under analysis. In particular, this scheme always selects the best relay among the available candidates, i.e., the node that experiences the most favorable channel conditions to the destination, and it resorts to an idealized selection process, so that all the potential candidates are immediately informed at no cost in terms of delay and overhead of how the hybrid ARQ phase will be organized, i.e., no contention is required and collisions among relays are avoided. Fig.~\ref{fig:rcSims_truBound} highlights that even in these ideal conditions the throughput improvement offered over CSMA never exceeds {15\%}. Such results, then, suggest that the potential of relaying strategies is severely hindered by some effects that arise in large carrier sense-based ad hoc networks.

\begin{figure}
\centering
\includegraphics[width=\figw]{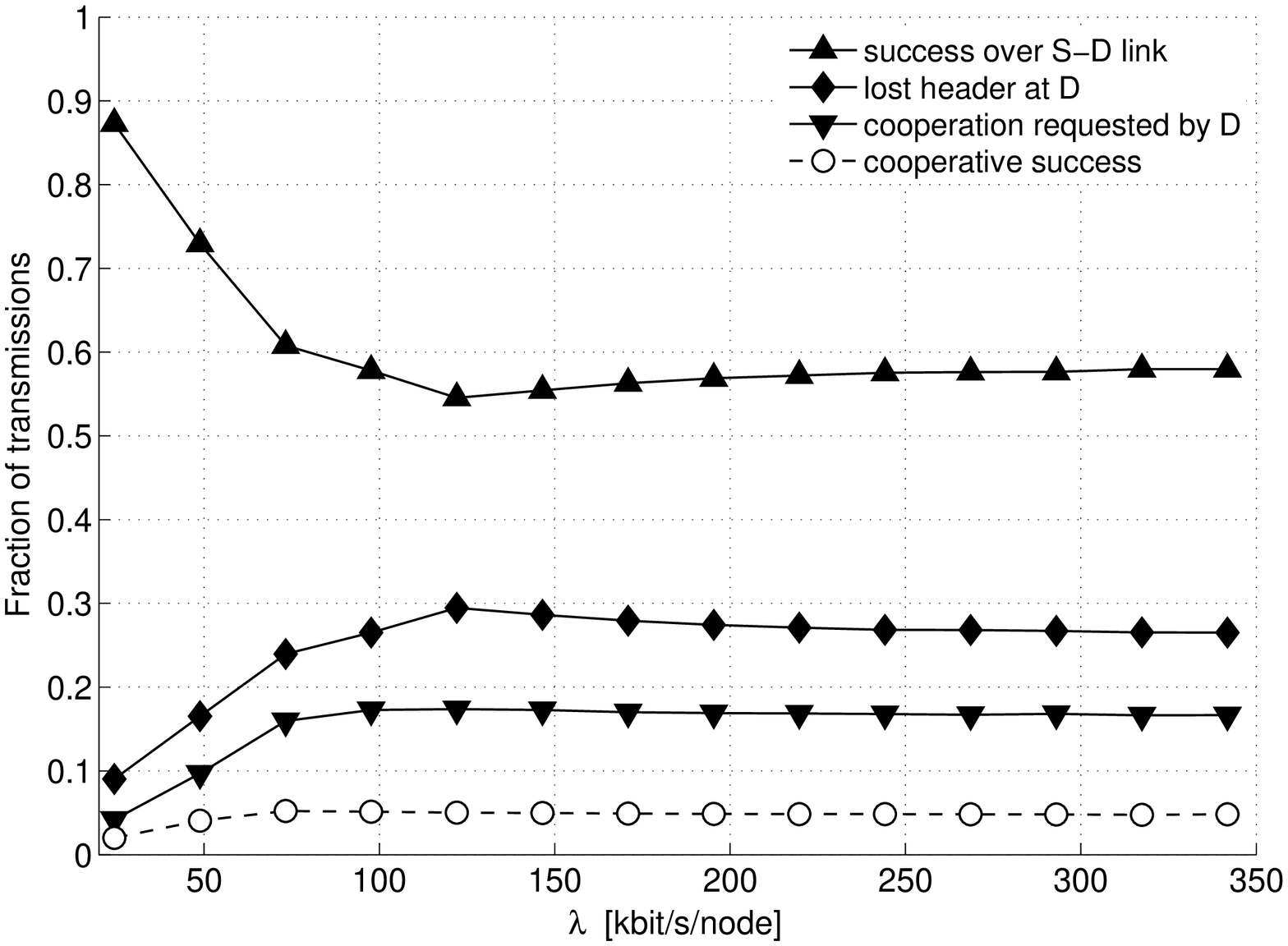}
\caption{Outcome of a single transmission vs nominal load. At the destination, three conditions
can occur: i) the transmission succeeds (\emph{success over S-D link}); ii) the destination is not able
to decode the header of the packet (\emph{lost header}); and iii) the header is decoded but the
payload is corrupted (\emph{cooperation requested}).}
\label{fig:rcSims_succHeaderCoop}
\end{figure}

In order to further investigate this aspect, we consider the outcome of a single transmission. As discussed in Section~\ref{sec:dharqDescription}, when a source sends a packet, three events can occur at the addressee: i) the transmission succeeds; ii) the destination is not able to decode the header of the packet; or iii) the header is decoded but the payload is corrupted. Cooperation can take place  only if condition iii) is met, since the destination needs to cache a corrupted version of the incoming data to take advantage of relayed retransmissions. Fig.~\ref{fig:rcSims_succHeaderCoop} shows the distribution of these events in our simulations. For low traffic, the success rate is rather high, and thus cooperation is seldom of use. Conversely, as $\lambda$ is raised, the probability of decoding the payload with a direct transmission decreases, due to the higher interference level. Nonetheless, the worse reception quality does not lead to a significant increase of the number of cooperative phases that may be expected. In fact, at saturation, transmission failures are mostly induced by the loss of the packet header, almost {30\%} of the communications, whereas relaying processes are curbed to less than {20\%} of the cases. In this perspective, it is important to observe that the incapability of retrieving packet headers does not primarily stem from channel impairments, as control frames are sent at a conservative rate $\rho_{ctrl}$, but rather from the completely distributed nature of the medium access policy under consideration. With CSMA, indeed, a destination may not perceive packets addressed to it since it is already synchronized to other ongoing communications, e.g., because of the hidden terminal problem. We remark that while this effect could not be taken into account in the simple scenarios analyzed in our theoretical studies, it plays a key role in larger and more congested networks and is fully captured by our simulation results.

\begin{figure}
\centering
\includegraphics[width=\figw]{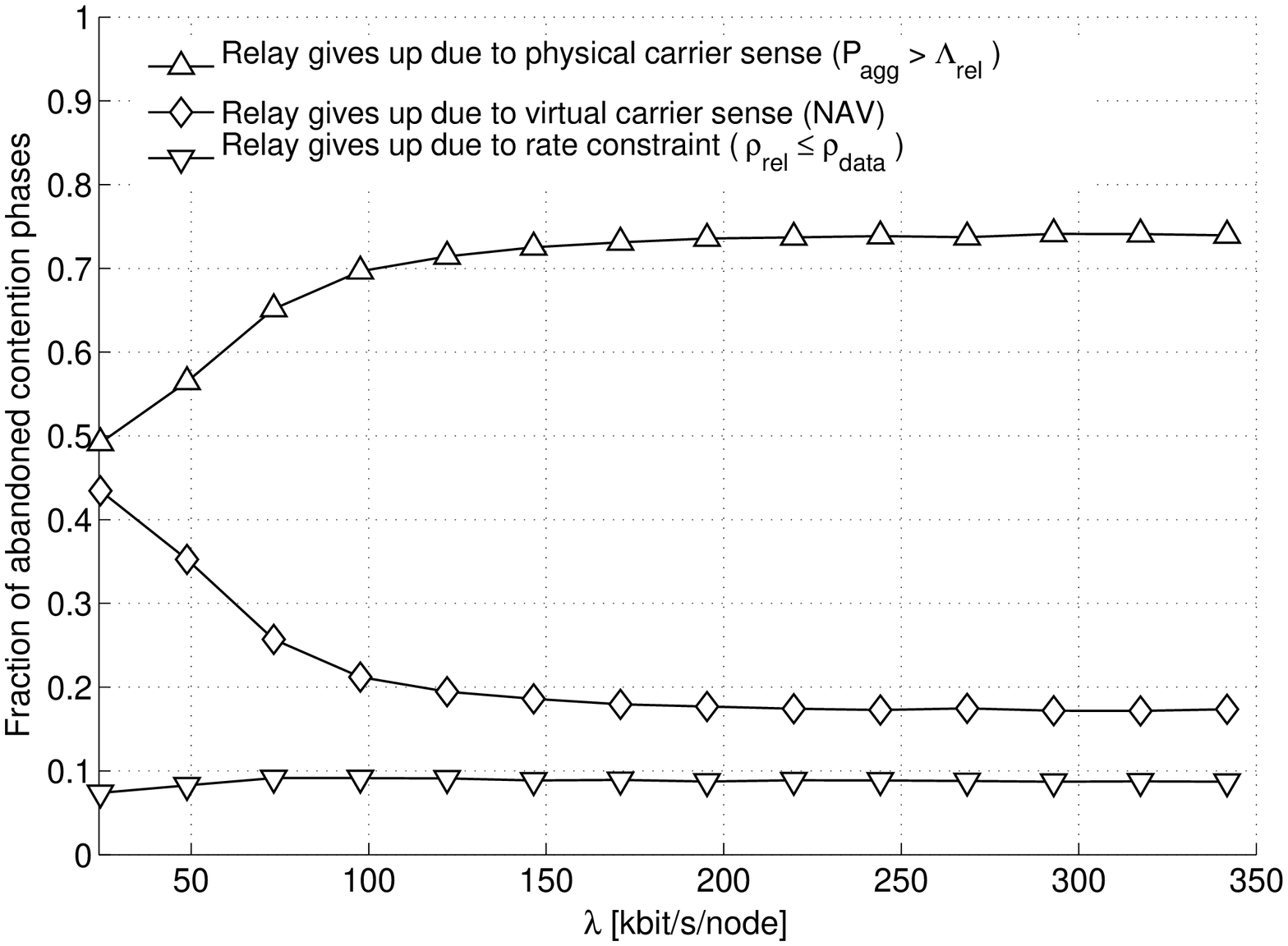}
\caption{Impact of the reasons that may lead a relay candidate to abandon a contention vs nominal load $\lambda$.}
\label{fig:rcSims_reasonsGiveUp}
\end{figure}

The reduced fraction of times in which a destination may successfully support relaying clearly represents a major limitation to the performance of DHARQ. However, Fig.~\ref{fig:rcSims_succHeaderCoop} also highlights that, even when requested, cooperation is successful only in one-fourth of the cases (dashed line). From this viewpoint, when a NACK is sent, the distributed hybrid ARQ mechanism that we propose may fail because of three reasons: i) no node has been able to decode the packet transmitted by the source; ii) some candidates are present, but all of them give up the contention; or iii) at least one terminal acts as relay, but the destination is still not able to decode the payload, e.g., because of a collision. Our simulations have shown that the dominant motive is by far factor ii), as nodes that abandon the election procedure are responsible for more than {60\%} of the non-performed cooperative phases. In turn, as reported in Fig.~\ref{fig:rcSims_reasonsGiveUp} against the nominal load, terminals refrain from relaying most likely because they perceive a power level on the medium which is above the carrier sense threshold $\Lambda_{rel}$ (80\% of the cases), whereas a much smaller number of withdrawals are due to the rate constraint described in Section~\ref{sec:dharqDescription}, or to virtual carrier-sense.\footnote{Packet headers contain information on the duration of the communication they initiate. Therefore, nodes that decode a header not addressed to them become aware that the channel will be busy for a given time, and suitably update their Network Allocation Vector (NAV) \cite{802.11}, deferring any channel access for the reserved period.} From this discussion we can once more infer that reactive cooperation is hampered by the inherent nature of carrier sensing, which often does not grant channel access to nodes that would be available for relaying.

\begin{figure}
\centering
\includegraphics[width=\figw]{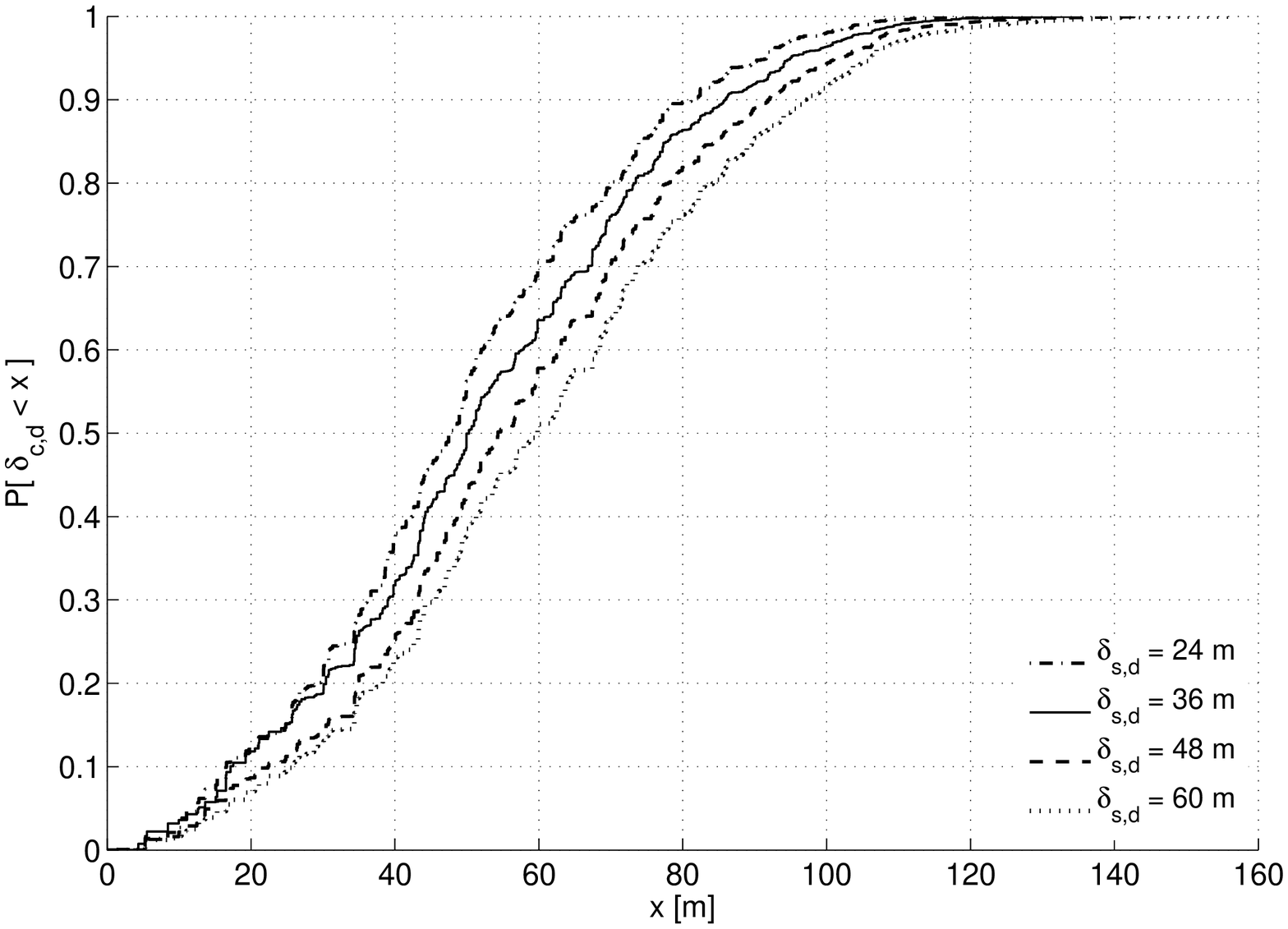}
\caption{Cumulative Distribution Function of the distance of relay candidates from the
destination. Different curves represent different source-destination distances.}
\label{fig:rcSims_cdfRelayPos}
\end{figure}

Finally, it is important to observe that not only does CSMA prevent a large share of cooperative phases, but also it reduces their effectiveness when they actually take place. As discussed in Section~\ref{sec:dharqAnalysis}, if an $\bm S$-$\bm D$ communication fails, interferers are most likely located toward $\bm D$ (see Fig.~\ref{fig:rcAnalysis_InterfDistrib}), since the area surrounding the source is protected by the carrier sense mechanism. This, in turn, tends to concentrate nodes that decode the payload, i.e., relay candidates, in the  proximity of $\bm S$. Thus, cooperators are likely to offer a limited (if any) advancement towards the destination. This has a negative effect on hybrid ARQ, since a higher path-loss induces longer retransmissions and therefore reduces the advantages of quick failure recovery offered by cooperation. We have delved into this aspect by performing a specific set of simulations. In particular, we considered topologies in which a source and a destination were located at the center of the network at distance $\delta_{s,d}$, whereas the rest of the terminals  were randomly distributed in the $300\,m \times 300\,m$ area. Fig.~\ref{fig:rcSims_cdfRelayPos} depicts the cumulative distribution function (CDF) of the distance from $\bm D$ of relay candidates, i.e., of nodes that decode the source's payload given that the reception at the destination fails, for values of $\delta_{s,d}$ ranging from 25 to 60 m. As a first remark, the plot confirms the bias in the positions of potential cooperators: on average, nodes that offer an advancement towards the destination with respect to the source are available only in {30\%} of the cases. Moreover, we notice that the smaller $\delta_{s,d}$, the higher the probability of finding relays that are in the proximity of the destination. This stems from the fact that, the closer $\bm S$ to $\bm D$, the more the area that carrier sense reserves for the ongoing communication extends over the destination's surroundings, making it possible for terminals in this region to participate in the relay contention. On the other hand, this favorable effect is not able to trigger significant improvements, since it facilitates relaying when cooperation tends to be triggered rarely anyway, due to the more favorable average channel conditions for links with reduced values of $\delta_{s,d}$.

\begin{figure}
\centering
\includegraphics[width=\figw]{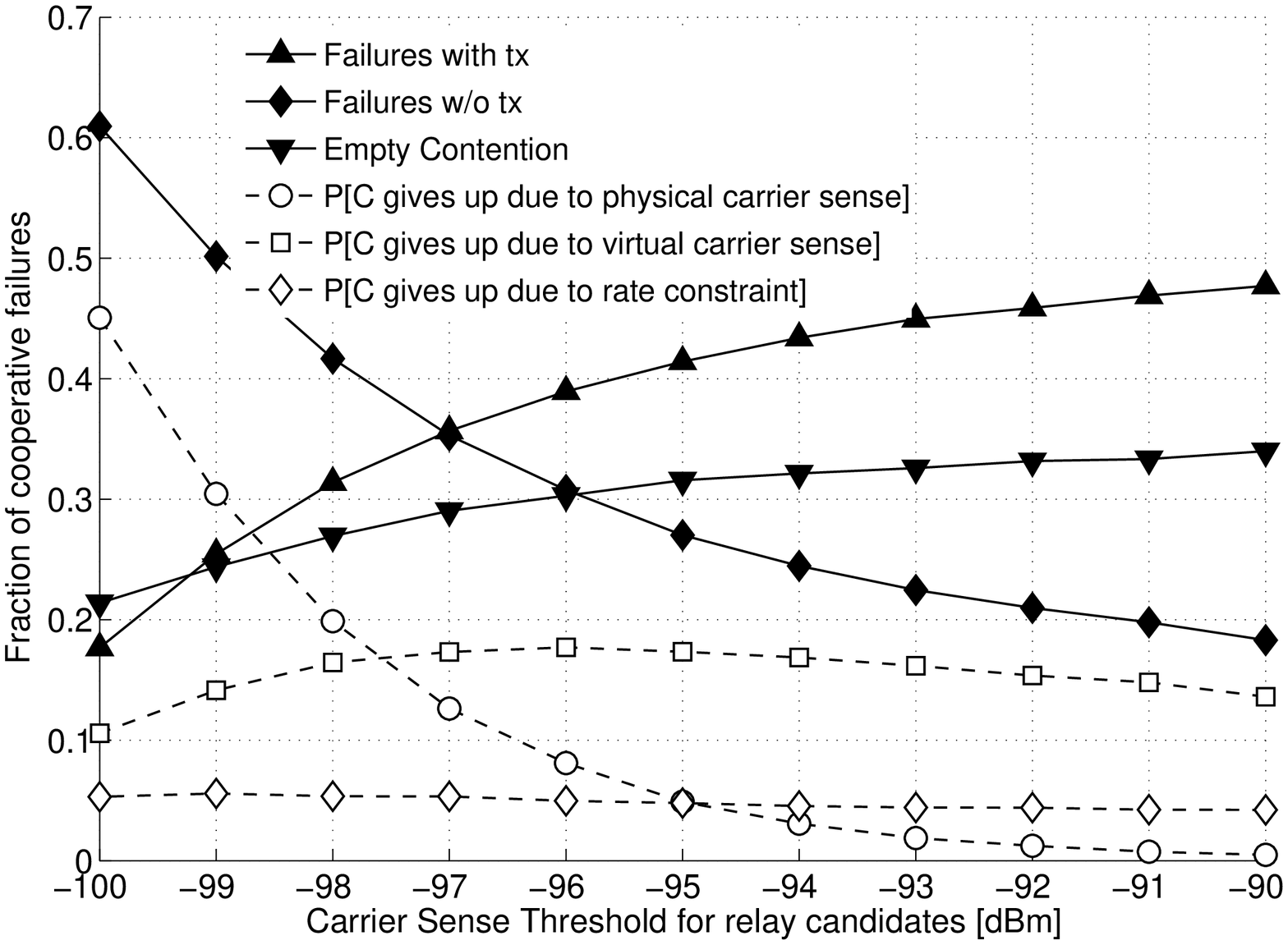}
\caption{Efficiency of cooperative phases for different values of the carrier sense threshold for
relay contentions.}
\label{fig:rcSims_failures}
\end{figure}

The results discussed so far have been obtained by setting the carrier sense threshold used during relay contention equal to the one used for basic channel access, i.e., $\Lambda_{rel} = \Lambda$. In order to increase the number of cooperative phases, one could think of a more aggressive approach, letting terminals act as cooperators even though they detect some activity on the channel. In our analysis, we have considered the effectiveness of such a strategy by testing DHARQ with different values of $\Lambda_{rel}$.
Let us first consider the results reported in Fig.~\ref{fig:rcSims_failures}. Here, the solid lines investigate why a relaying request (i.e., a NACK packet) is not followed by a cooperative success. As discussed earlier, such a condition may happen either because no node has decoded the payload sent by the source (\emph{empty contention}), or because all the candidates give up the contention (\emph{failure w/o tx}), or because at least one relay actually transmits but the destination is not able to decode the payload, likely due to a collision (\emph{failure with tx}). As $\Lambda_{rel}$ is increased, nodes that have cached the packet sent by the source tend to become more aggressive, and more cooperative phases take place, as shown by the fact that the impact of failures without transmissions significantly decreases, passing from {60\%} to as low as {20\%}. This behavior is further confirmed by the dashed lines in the plot, which report the reasons that lead nodes to refrain from cooperating and sum up to the \emph{failure w/o tx} curve: a higher carrier sense threshold prevents relay candidates from withdrawing from the contention. However, this approach has two main consequences. First of all, the number of collisions among relays increases, since candidate nodes may not overhear each other's attempt and may access the channel simultaneously. We stress that while on the one hand concurrent relay transmissions are likely to significantly worsen the effectiveness of cooperation, on the other hand they also represent a waste of network resources in terms of energy and bandwidth. Moreover, a higher number of relayed packets tends to raise the overall interference, and thus to lower the average decoding probability in the system. Not only does this affect reception at destination nodes, but also it reduces the number of potential relays that are able to successfully retrieve the packet sent by the source. Fig.~\ref{fig:rcSims_failures} confirms this observation, showing how the number of times in which no terminal has been able to decode the payload, i.e., empty contentions, increases with $\Lambda_{rel}$.

\begin{figure}
\centering
\includegraphics[width=\figw]{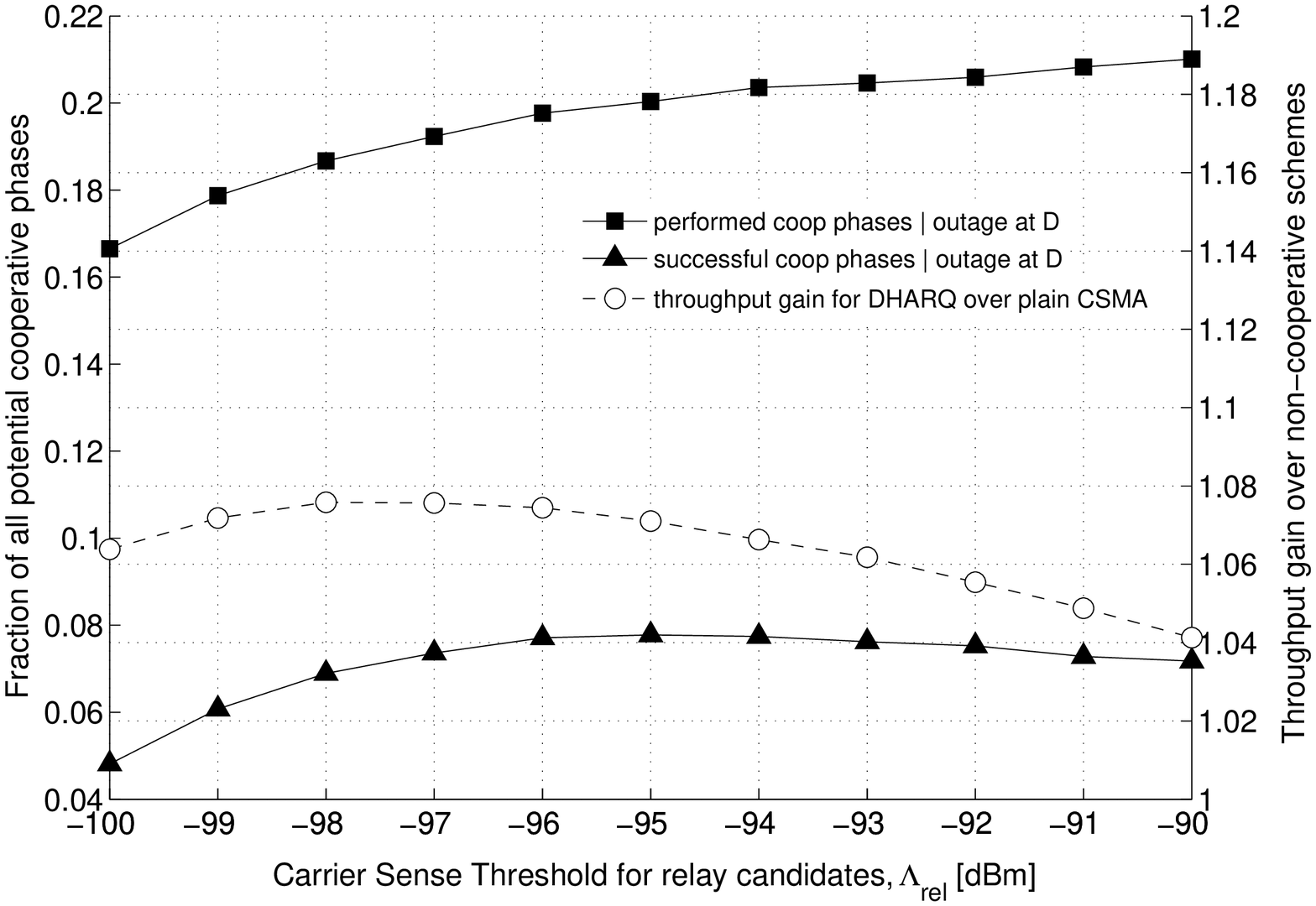}
\caption{Impact of cooperative phases and aggregate throughput gain with respect to CSMA for different values of the carrier sense threshold for relay contentions.}
\label{fig:rcSims_truVsCsThreshold}
\end{figure}

The aforementioned tradeoffs reflect at a system level, as reported in Fig.~\ref{fig:rcSims_truVsCsThreshold}. When $\Lambda_{rel}$ is raised, more relays tend to actually perform retransmissions (left axis, square-marked curve). This initially leads to an increase in the fraction of successful cooperative phases as well (left axis, triangle-marked curve), which pass from the 5\% that characterized the reference implementation discussed in Fig.~\ref{fig:rcSims_succHeaderCoop} to up to 8\% of the overall communications. However, as soon as the carrier sense threshold goes beyond a certain limit (around $-97$ dBm in our settings), the drawbacks induced by collisions and by the higher level of interference prevail, worsening the overall efficiency of the cooperative mechanism. As a consequence, a similar trend is followed by the throughput gain of DHARQ, defined as the ratio of the aggregate throughput of the cooperative solution over the same quantity achieved by plain CSMA. As shown in the plot (right axis, white markers), such an overall improvement metric plummets when relays become too aggressive. These results suggest that the optimal working point for the system can be achieved for a value of $\Lambda_{rel}$ slightly higher than the one used for basic channel access. Nevertheless, the benefits that can be obtained are limited, as the gains over the benchmark protocol never exceed {10\%}.

In conclusion, the simulation studies reported in this paper have thoroughly highlighted a detrimental effect of CSMA on reactive cooperation, induced both by the bias in the spatial distribution of relay nodes discussed in Section~\ref{sec:dharqAnalysis} and by other simple yet unavoidable networking mechanisms that would affect any implementation of a real-word cooperative scheme in a carrier sense-based environment.

\section{Conclusions} \label{sec:conclusions}

This paper represents Part~I of a two-part work that thoroughly discusses the issues arising when cooperative paradigms are implemented
in non-infrastructured wireless ad hoc networks with carrier sense-based channel contention. The present paper has focused on \emph{reactive} solutions, which trigger relaying only in the event of a packet delivery failure over the direct source-destination link. In particular, by means of both analysis and simulations, we showed that the CSMA policy biases the spatial distribution of the available cooperators, in such a way that the theoretical performance gain granted by relaying is significantly reduced. Moreover, the presence of interferers in the proximity of cooperating nodes diminishes diversity at their intended receiver, further hampering the effectiveness of distributed hybrid ARQ. Finally, the impact of several practical issues, such as synchronization and the hidden terminal problem has been studied. The companion paper \cite{part2} tackles similar issues when \emph{proactive} cooperative schemes are implemented, in which the source node and the relays take advantage of channel state information to preemptively coordinate with the aim of maximizing the transmission bitrate. 

\appendix[]
\subsection{Derivation of $\mathcal O^{t_{i}}(\bm p)$}
\label{appendixDHARQ_interf}

With reference to the scenario described in Section~\ref{sec:dharqAnalysis}, we 
shall compute the probability that the interfering terminal $\bm I$
induces an outage at the destination given that both $\bm S$ and $\bm I$ have
been granted access to the medium, i.e., $\mathcal O^{t_{i}}(\bm p)$.
Conditioning on the power received over the $\bm S$-$\bm D$ link, we get:
\begin{equation}
\mathcal O^{t_{i}}(\bm p) \!=\!\! \int_{0}^{+\infty} \!\!\!\textrm{Pr} \! \left\{\mathcal
L^{t_{i}}(\bm \eta) \! <  \! L \,|\, \eta_{s,d}\right \} f(\eta_{s,d}, P
\delta^{-\alpha}_{s,d}) \; d\eta_{s,d}
\label{eq:conditionalOutage}
\end{equation}
where $f(\eta, a) = e^{ - \frac{\eta}{a}} /a$, and $\mathcal L^{t_{i}}(\bm \eta)$, represented by (\ref{eq:decodedInfoBits_ti}), is the number of information bits retrieved at $\bm D$ for a topology $\bm p = \{p_s, p_d, p_i\}$ and conditioned on the birth time $t_i$ of $\bm I$'s transmission. In order to solve (\ref{eq:conditionalOutage}), we identify three disjoint regions whose union spans the whole integration domain: $\mathcal R_{1}
= [0, \eta^{*})$, $\mathcal R_{2} = [\eta^{*}, \bar{\eta})$, and $\mathcal R_{3} =
[\bar{\eta}, +\infty)$, so that for $\eta_{s,d} \in \mathcal R_{1}$ an outage event is induced
by the sole noise power, regardless of the values of $t_{i}$ and $\eta_{i,d}$, whereas for $\eta_{s,d}
\in \mathcal R_{3}$ the capacity of the $\bm S$-$\bm D$ channel is sufficient to support a
successful communication during the interference-free interval of duration $T - |t_i|$. The values
of $\eta^{*}$ and $\bar{\eta}$ can be determined imposing the conditions $\mathcal
C(\eta^{*}/N) = L$, and $|t_{i}| \, \mathcal C(\bar{\eta}/N) = L$, obtaining:
$\eta^{*} = N\, (2^{\rho_{data}/B} - 1)$ and $\bar{\eta} = N\, (2^{L/(B \, |t_i|)} - 1)$,
respectively. Taking advantage of this subdivision, it immediately follows that the integration
over $\mathcal R_{3}$ returns zero, while the integrand over $\mathcal R_{1}$ simplifies to
$f(\eta_{s,d}, P\delta^{-\alpha}_{s,d})$.
Therefore, we get:
\begin{equation}
\mathcal O^{t_{i}}(\bm p) = 1 - e^{-\frac{\eta^{*}}{P\delta_{s,d}^{-\alpha}}}
\,\,+ \int_{\eta^{*}}^{\bar{\eta}} e^{-\frac{\xi(\eta_{s,d}, t_{i})}{P\delta^{-\alpha}_{i,d}}}
f(\eta_{s,d}, P \delta^{-\alpha}_{s,d})
\; d\eta_{s,d} \,,
\label{eq:outageProb_ti}
\end{equation}
where 
\begin{equation}
\xi(\eta_{s,d}, t_{i}) = \frac{\eta_{s,d}} {\left( \frac{2^{L / (B\, |t_i|)}}  
{1 + \frac{\eta_{s,d}}{N}} \right)^{\frac{|t_i|}{T-|t_i|}} -1 } - N 
\label{eq:xi}
\end{equation}
is obtained by solving $\mathcal L^{t_{i}}(\bm \eta) = L$ with respect to $\eta_{i,d}$.\footnote{In greater detail, recalling (\ref{eq:decodedInfoBits_ti}) and (\ref{eq:decodedInfoBits}), $\xi(\eta_{s,d}, t_{i})$ can be computed by finding the value of $\eta_{i,d}$ that satisfies $|t_i| \, B \log_2\left( 1 + \eta_{s,d}/N \right) + (T - |t_i|) \, B \log_2 \left( 1 + \eta_{s,d}/(N + \eta_{i,d}) \right) = L$.} The value of $\mathcal O^{t_i}(\bm p)$ for any topology $\bm p$ and for any value of $t_i$ can be computed by numerical evaluation of (\ref{eq:outageProb_ti}).

\subsection{Derivation of $\mathcal H^{p_i, t_{i}}_+(\bm p')$}
\label{appendixDHARQ_coop}

Let us refer to the topology described in Section~\ref{sec:dharqAnalysis}, and let $\bm \eta' = \{
\eta_{s,c}, \eta_{i,c} \}$ be the vector of the powers received at $\bm C$ from $\bm S$ and $\bm
I$, respectively. Furthermore, let $\mathcal L_c^{t_i}(\bm \eta')$ be the number of
information bits sent by the source that are decoded at such node conditioned on the start time
$t_i$ of the interfering communication.\footnote{$\mathcal L_c^{t_i}(\bm \eta')$ is promptly given
by (\ref{eq:decodedInfoBits}), replacing $\eta_{s,d}$ and $\eta_{i,d}$ with $\eta_{s,c}$
and $\eta_{i,c}$ respectively.} According to the discussion in Section~\ref{sec:dharqAnalysis},
the probability that $\bm C$ is available for cooperation, under the assumption that $t_i >0$, can
be written as:
\begin{equation}
 \mathcal H^{p_i, t_i}_+ (\bm p') = \textrm{Pr}\{ \mathcal L_c^{t_i}(\bm \eta') > L
\;, \; \eta_{i,c} + N < \Lambda  \} \;.
\label{eq:h+_generica}
\end{equation}
Let now $\mathcal H^{p_i, t_i,\eta_{s,c}}_+$ represent (\ref{eq:h+_generica}) conditioned on
$\eta_{s,c}$, so that 
\begin{equation}
 \mathcal H^{p_i, t_i}_+ (\bm p') \!=\!\! \int_0^{+\infty} \!\! \mathcal H^{p_i,
t_i, \eta_{s,c}}_+ (\bm p') \, f(\eta_{s,c}, P \delta_{s,c}^{-\alpha}) \; d\eta_{s,c}.
\label{eq:h+_conditioned}
\end{equation}

In order to solve this equation, we split the integration domain in three disjoint
regions corresponding to those identified in Appendix~\ref{appendixDHARQ_interf}: $\mathcal R_{1} =
[0, \eta^*)$, $\mathcal R_{2} = [\eta^*, \bar \eta)$, and $\mathcal R_{3} = [\bar \eta, +\infty)$.
When $\eta_{s,c}\in \mathcal R_1$, the sole noise induces an outage at $\bm C$ so that the region
does not contribute to the overall computation. On the other hand, when $\eta_{s,c} \in \mathcal
R_3$, decoding is ensured regardless of the interfering power. It follows that in this case the
integrand reduces to $\textrm{Pr}\{\eta_{i,c} + N < \Lambda\} \, f(\eta_{s,c},P
\delta_{s,c}^{-\alpha})$, and (\ref{eq:h+_conditioned})  restricted to $\mathcal R_3$
returns:
\begin{equation}
 \left( 1 + e^{- \frac{\Lambda - N}{P \delta_{i,c}^{-\alpha}}} \right) \, e^{-\frac{\bar \eta}{P
\delta_{s,c}^{-\alpha}}}\;.
\label{eq:h+_firstTerm}
\end{equation}

Finally, if $\eta_{s,c} \in \mathcal R_2$, the level of interference determines both the decoding
probability and the probability that $\bm C$ senses the medium idle. Therefore, solving
$\mathcal L_c^{t_i}(\bm \eta') > L$ with respect to $\eta_{i,c}$, we can write the
contribution of
$\mathcal R_2$ to $ \mathcal H^{p_i, t_i}_+ (\bm p')$ as:
\begin{equation}
 \int_{\eta^*}^{\bar \eta} \textrm{Pr} \left\{ \eta_{i,c} < \xi(\eta_{s,c}, t_i) \;, \;
 \eta_{i,c} < \Lambda - N \,|\, \eta_{s,c} \right\} \, f(\eta_{s,c},P \delta_{s,c}^{-\alpha}) \;d\eta_{s,c} \;.
\label{eq:h+_secondTerm}
\end{equation}

$\xi(\eta_{s,c}, t_i)$, defined in (\ref{eq:xi}), has a zero for $\eta_{s,c} = \eta^*$, is
monotonically increasing in the interval $[\eta^*, \bar \eta)$, and exhibits a vertical asymptote
at $\bar \eta$. Let us now define a value $\tilde \eta$ so that $\xi(\tilde \eta, t_i) =
\Lambda - N$. From the aforementioned properties, we can infer that $\min\{\xi(\eta_{s,c},t_i),
\Lambda - N\}$ is equal to $\xi(\eta_{s,c},t_i)$ for $\eta_{s,c} \in [\eta^*, \tilde\eta )$,
while for $\eta_{s,c} \in [\tilde\eta, \bar \eta)$ the more stringent requirement is given by
$\Lambda - N$. Therefore, (\ref{eq:h+_secondTerm}) can be rewritten as:
\begin{equation}
 \int_{\eta^*}^{\tilde \eta} \left( 1 - e^{-\frac{\xi(\eta_{s,c},t_i)}{P \delta_{i,c}^{-\alpha}}}
\right) f(\eta_{s,c}, P \delta_{s,c}^{-\alpha}) \; d\eta_{s,c} +
 \int_{\tilde \eta}^{\bar \eta} \left( 1 - e^{-\frac{\Lambda - N}{P
\delta_{i,c}^{-\alpha}}} \right)  f(\eta_{s,c}, P \delta_{s,c}^{-\alpha}) \; d\eta_{s,c} \;.
\label{eq:h+_expanded}
\end{equation}

Finally, combining the results of (\ref{eq:h+_firstTerm}) and (\ref{eq:h+_expanded}), we obtain:
\begin{equation}
 \mathcal H^{p_i, t_i}_+ (\bm p') = e^{-\frac{\tilde \eta}{P
\delta_{s,c}^{-\alpha}}} \, \left( 1 - e^{-\frac{\Lambda - N}{P \delta_{i,c}^{-\alpha}}} \right)
+ \int_{\eta^*}^{\tilde \eta} \!\! \left( 1 - e^{-\frac{\xi(\eta_{s,c},t_i)}{P
\delta_{i,c}^{-\alpha}}} \right) \! f(\eta_{s,c}, P \delta_{s,c}^{-\alpha})
\;d\eta_{s,c}. \nonumber 
\end{equation}
Given the complexity of the involved functions, it is not possible to find a closed form for the
considered probabilities. However, $\tilde \eta$ and consequently $\mathcal H^{p_i, t_i}_+(\bm p')$ can be
determined by means of numerical evaluation for any topology and for any value of $t_i$.

\bibliographystyle{IEEEtran}
\bibliography{IEEEabrv,TWbib}

\end{document}